\documentclass[aps,preprint]{revtex4-1} 
\usepackage{graphicx}
\usepackage{graphics}
\usepackage{amsmath}

\newcommand{\half}{\frac{1}{2}}

\newcommand{\eps}{\epsilon}

\newcommand\refeq[1]{(\ref{#1})}
\newcommand\reffig[1]{Figure~\ref{#1}}
\newcommand\of[1]{\left( #1 \right)}
\newcommand\sqof[1]{\left[ #1 \right]}
\newcommand\vecbf[1]{{\bf #1}}
\newcommand\meskip{\hbox{\hskip 1cm}}

\begin{document}

\title{The Motion of a Pair of Charged Particles}

\author{J. Franklin}
\email{jfrankli@reed.edu}
\author{C. LaMont}

\affiliation{Department of Physics, Reed College, Portland, Oregon 97202,
USA}

\begin{abstract}
We re-visit the problem of two (oppositely) charged particles interacting electromagnetically in one dimension with retarded potentials and no radiation reaction.  The specific quantitative result of interest is the time it takes for the particles to fall in towards one another.  Starting with the non-relativistic form, we answer this question while adding layers of complexity until we arrive at the full relativistic delay differential equation that governs this problem.  That case can be solved using the Synge method, which we describe and discuss. 
\end{abstract}

\maketitle

\section{Introduction}

The fields that solve Maxwell's equations for a charged point particle traveling along a trajectory $\vecbf w(t)$ are given by (see~\cite{Griffiths}, for example):
\begin{equation}\label{EBpoints}
\begin{aligned}
\vecbf E(\vecbf r, t) &= \frac{q}{4 \, \pi \, \epsilon_0} \, \frac{R}{\of{\vecbf R \cdot \vecbf u}^3} \, \left[ (c^2 - v^2) \, \vecbf u + \vecbf R \times \of{\vecbf u \times \vecbf a} \right] \\
\vecbf B(\vecbf r, t) &= \frac{1}{c} \, \hat{\vecbf R} \times \vecbf E(\vecbf r, t)
\end{aligned}
\end{equation}
for $\vecbf R \equiv \vecbf r - \vecbf w$,  $\vecbf v = \dot{\vecbf w}$, $\vecbf a = \ddot{\vecbf w}$, $\vecbf u \equiv c \, \hat{\vecbf R} - \vecbf v$ and 
where $\vecbf w$ and all of its time-derivatives are evaluated at the retarded time, $t_r$, defined implicitly by:
\begin{equation}\label{implicitdef}
c \, (t - t_r) = \| \vecbf r - \vecbf w(t_r)\|.
\end{equation}
We have included only the causally relevant retarded time contribution to the field at point $\vecbf r$, time $t$.

A particle traveling under the influence of these fields moves according to the relativistic equation of motion:
\begin{equation}\label{reom}
\frac{d}{dt} \left[ \frac{m \, \vecbf v}{\sqrt{1 - \frac{v^2}{c^2}}}\right] = q \, \vecbf E + q \, \vecbf v \times \vecbf B,
\end{equation}
which is Newton's second law with relativistic momentum and electromagnetic forces -- we have omitted the radiation reaction force on the right.  The combination~\refeq{EBpoints}, ~\refeq{implicitdef}, ~\refeq{reom}, taken for a pair of charged particles (where each particle is dynamically influenced by the field of the other) defines the ``Synge" version of the problem of charged particle motion (as in~\cite{Synge}).  That formulation omits radiation reaction in part because such effective forces do not come from the usual starting action for this two-particle relativistic problem (the Fokker-Schwarzschild-Tetrode one, see~\cite{Barut}).  Variants include combinations of half-advanced and half-retarded interaction, where for example, circular orbits are available~\cite{Schild}.

Our goal in this paper is to study the one-dimensional motion of two interacting charged particles progressively, starting from the non-relativistic Coulomb interaction, then introducing relativistic dynamics, and finally, the full delay differential equation implied by the Synge formulation above.   We will take, as our model setup, a pair of equal and opposite charges, $\pm q$, with identical mass $m$ sitting on the $\hat{\vecbf x}$ axis.  At time $t = 0$, the charges are separated by a distance $2 \, d$ and at rest -- if we set the charges symmetrically about zero, then we can describe the motion of the positive charge using $x(t)$, and the negative charge will be at $-x(t)$.  The equation of motion for the positive particle is then:
\begin{equation}
\begin{aligned}
\frac{d}{dt} \, \sqof{ \frac{m \, \dot x(t)}{\sqrt{1 - \frac{\dot x(t)^2}{c^2}}} }
&= - \frac{q^2}{4 \, \pi \, \eps_0} \, \frac{1}{\of{ x(t) + x(t_r)}^2 } \, \frac{c - \dot x(t_r)}{c + \dot x(t_r)} \\
c \, \of{t - t_r} &= |x(t) + x(t_r)|.
\end{aligned}
\end{equation}
The problem will be to find the time it takes the pair of particles to halve their initial separation distance -- we are looking for $t^*$ such that $x(t^*) = d/2$.   In the form above, it is clear that finding $x(t^*)$ requires knowledge of the particle motion for all times $t < 0$ (since at $t = 0$, the right-hand-side of the equation of motion requires us to know the location of the particle at $t_r < 0$).  That is more initial data than is ordinarily required in classical mechanics (where the initial position and velocity suffice to determine $x(t^*)$).  It is not immediately clear how we should specify this initial data -- we don't have control of the particles for all time prior to $t = 0$, so at best we can only approximately specify the past history.

We will peel away the complicating elements of the full problem by studying the answer to our question (``what is $t^*$?") in approximations to the full problem, and comparing those solutions.   The specific regimes of interest to us will be:  1.  non-relativistic infall, 2.\  relativistic infall with specified initial conditions for all times prior to $t = 0$ -- this case includes relativistic dynamics and the full relativistic field both without retardation and with retardation, and 3.\ full relativistic dynamics with retardation and no assumptions about the motion of the particles prior to $t = 0$.   In this final case, we will use a method developed by Synge in $1940$ (\cite{Synge}), and re-developed independently in~\cite{Colin}.
Again, the goal is to compare these various regimes against each other, to see how the physical mechanisms and predictions change in each regime.  The application of the Synge iteration technique is new to this particular form of the problem, although it was successfully applied to opposite charges traveling in one dimension with both retarded and advanced potentials~\cite{Andersen}.

\section{Non-Relativistic Infall}

In the simplest case, where the particles are not traveling too fast, and are relatively close together, we can use the Coulomb force.  The force on the positive charge, due to the negative one is:
\begin{equation}
F = -\frac{q^2}{4 \, \pi \, \eps_0 \of{x(t) - (-x(t))}^2},
\end{equation}
and then the equation of motion is just
\begin{equation}
\ddot x(t) = -\frac{q^2}{4 \, \pi \, \eps_0 \, m\, (2\, x(t))^2}.
\end{equation}
We'll introduce dimensionless quantities, in preparation for the numerical work to come -- the convention will be that lower case variables have dimension, and their upper case forms are dimensionless.  So set $x = x_0 \, X$ and $t = (x_0/c) \, T$, then the equation of motion is
\begin{equation}\label{CCF}
\frac{d^2 X}{d T^2} =-\underbrace{\frac{ \frac{q^2}{4 \, \pi \, \eps_0 \, x_0}}{m \, c^2}}_{\equiv \alpha} \, \frac{1}{4 \, X^2},
\end{equation}
where the dimensionless parameter $\alpha$ is the ratio of the characteristic electrostatic potential energy to the relativistic rest energy.

We can multiply both sides of~\refeq{CCF} by $\frac{d X}{d T} \equiv X'$ and integrate, recovering the usual conservation of energy for particle motion in E\&M:
\begin{equation}\label{consCCF}
\half \of{X'}^2 = \frac{\alpha}{4 \, X} + E_x.
\end{equation}
The constant $E_x$ can be set using the initial conditions: $X(0) = D \equiv d/x_0$ and $X'(0) = 0$ gives $E_x = -\frac{\alpha}{4 \, D}$.  Now taking the square root of both sides of \refeq{consCCF} (choose the negative root for infall), we can set up an integral to find $T^*$:
\begin{equation}
-\frac{1}{\sqrt{\alpha}} \, \int_{D}^{D/2} \frac{dX}{\sqrt{\frac{1}{X} - \frac{1}{D}} } = \int_0^{T^*} dT = T^*,
\end{equation}
or, performing the integration:
\begin{equation}\label{nonTstar}
T^* = \sqrt{\frac{1}{8 \, \alpha}} \, \of{2 + \pi} \, D^{3/2}.
\end{equation}

\section{Relativistic Infall}

The full point-source fields of E\&M depend on both the position and velocity of the particle producing the field. In addition, the evaluation of the field at any location $x$ must take into account the amount of time it took for the field information to come from the source particle location at $x'$.  That information travels at $c$, so the time of flight is defined by
\begin{equation}
(t - t_r) = |x - x'|/c.
\end{equation}
Since we are studying moving particles, $x(t)$ (the location of the charge $q$) and $x'(t)$ (the location of the charge $-q$) are both functions of time, and the retarded time definition becomes an implicit one:
\begin{equation}
c \, (t - t_r) = |x(t) - x'(t_r)|.
\end{equation}

For our symmetric setup, where the particles lie on a line, the force on the particle at $x(t)$ due to the particle at $-x(t)$ is given by (see~\cite{Griffiths}, Problem 10.18):
\begin{equation}\label{fullforce}
F=- \frac{q^2}{4 \, \pi \, \eps_0} \, \frac{1}{\of{ x(t) + x(t_r)}^2 } \, \frac{c - \dot x(t_r)}{c + \dot x(t_r)}.
\end{equation}

In addition to using the full form of the fields coming from the Li\'enard-Wiechert potentials, we must also use the relativistic form of Newton's second law:
\begin{equation}
\frac{d}{dt} \, \sqof{ \frac{m \, \dot x(t)}{\sqrt{1 - \frac{\dot x(t)^2}{c^2}}} }= F.
\end{equation}
Putting these together with the definition of retarded time we have the following nonlinear delay differential equation for the motion of the positive charge located at $x(t)$:
\begin{equation}\label{relforce}
\begin{aligned}
\frac{d}{dt} \, \sqof{ \frac{m \, \dot x(t)}{\sqrt{1 - \frac{\dot x(t)^2}{c^2}}} }
&= - \frac{q^2}{4 \, \pi \, \eps_0} \, \frac{1}{\of{ x(t) + x(t_r)}^2 } \, \frac{c - \dot x(t_r)}{c + \dot x(t_r)} \\
c \, \of{t - t_r} &= |x(t) + x(t_r)|.
\end{aligned}
\end{equation}
We'll record the full problem in our dimensionless variables,
\begin{equation}\label{fullproblemdless}
\begin{aligned}
\frac{d}{dT} \, \sqof{ \frac{X'(T)}{\sqrt{1 - X'(T)^2}}}& = -\frac{\alpha}{\of{X(T) + X(T_r)}^2} \, \frac{1 - X'(T_r)}{1 + X'(T_r)} \\
T - T_r &= X(T) + X(T_r).
\end{aligned}
\end{equation}

\subsection{No Retardation}

Our first approximation will be to take $t_r = t$ in the evaluation of the force~\refeq{fullforce}, and this amounts to assuming that the particles are so close together that the time-of-flight for light is essentially zero.  Here, then, the problem reduces to:
\begin{equation}
\frac{d}{dT} \, \sqof{ \frac{X'(T)}{\sqrt{1 - X'(T)^2}}} = -\frac{\alpha}{4 \, X(T)^2} \, \frac{1 - X'(T)}{1 + X'(T)},
\end{equation}
which can be integrated once.  Rather than do that, we'll use this as an opportunity to define the numerical method we will use for the rest of the investigations in this paper.

We'll start by rendering the second order differential equation into a pair of first order ones -- define the dimensionless relativistic momentum $P(T) \equiv X'(T)/\sqrt{1 - X'(T)^2}$, and then we have:
\begin{equation}\label{XP}
\begin{aligned}
\frac{d X}{d T} &= \frac{P}{\sqrt{1 + P^2}} \\
\frac{d P}{d T} &=  -\frac{\alpha}{4 \, X^2} \, \frac{1 -  \frac{P}{\sqrt{1 + P^2}} }{1 +  \frac{P}{\sqrt{1 + P^2}} } = -\frac{\alpha}{4 \, X^2} \, \sqof{1 + 2 \, P \of{P - \sqrt{1 + P^2}}}.
\end{aligned}
\end{equation}

To approximate the solution to this pair of equations numerically, we define a grid in (dimensionless) time: $T_j \equiv j \, \Delta T$ for constant spacing $\Delta T$.  Now assume $X(T_j) \equiv X_j$ and $P(T_j) \equiv  P_j$ are the projections of the true solution onto the grid.  The derivatives can be approximated using finite differences:
\begin{equation}
X'(T_j) \approx \frac{X_{j+1} - X_{j-1}}{2 \, \Delta T}  \meskip P'(T_j) \approx \frac{P_{j+1} - P_{j-1}}{2 \, \Delta T}.
\end{equation}
Using these in~\refeq{XP} we can define a recursive update scheme for $X_j$ and $P_j$:
\begin{equation}\label{noretiter}
\begin{aligned}
X_{j+1} &= X_{j-1} + 2 \, \Delta T \, \frac{P_j}{\sqrt{1 + P_j^2}} \\
P_{j+1} &= P_{j-1}  -\frac{\alpha \, \Delta T}{2 \, X_j^2} \, \sqof{1 + 2 \, P_j \of{P_j - \sqrt{1 + P_j^2}}}.
\end{aligned}
\end{equation}
All we need to get this numerical procedure going is $X_{-1}$ and $P_{-1}$, the position and momentum of the positive charge just prior to $T = 0$ -- let's take those to be $X_{-1} = D$ and $P_{-1} = 0$, i.e.\ they will take on their values at $T = 0$.

\begin{figure}[htbp] 
   \centering
   \includegraphics[width=4in]{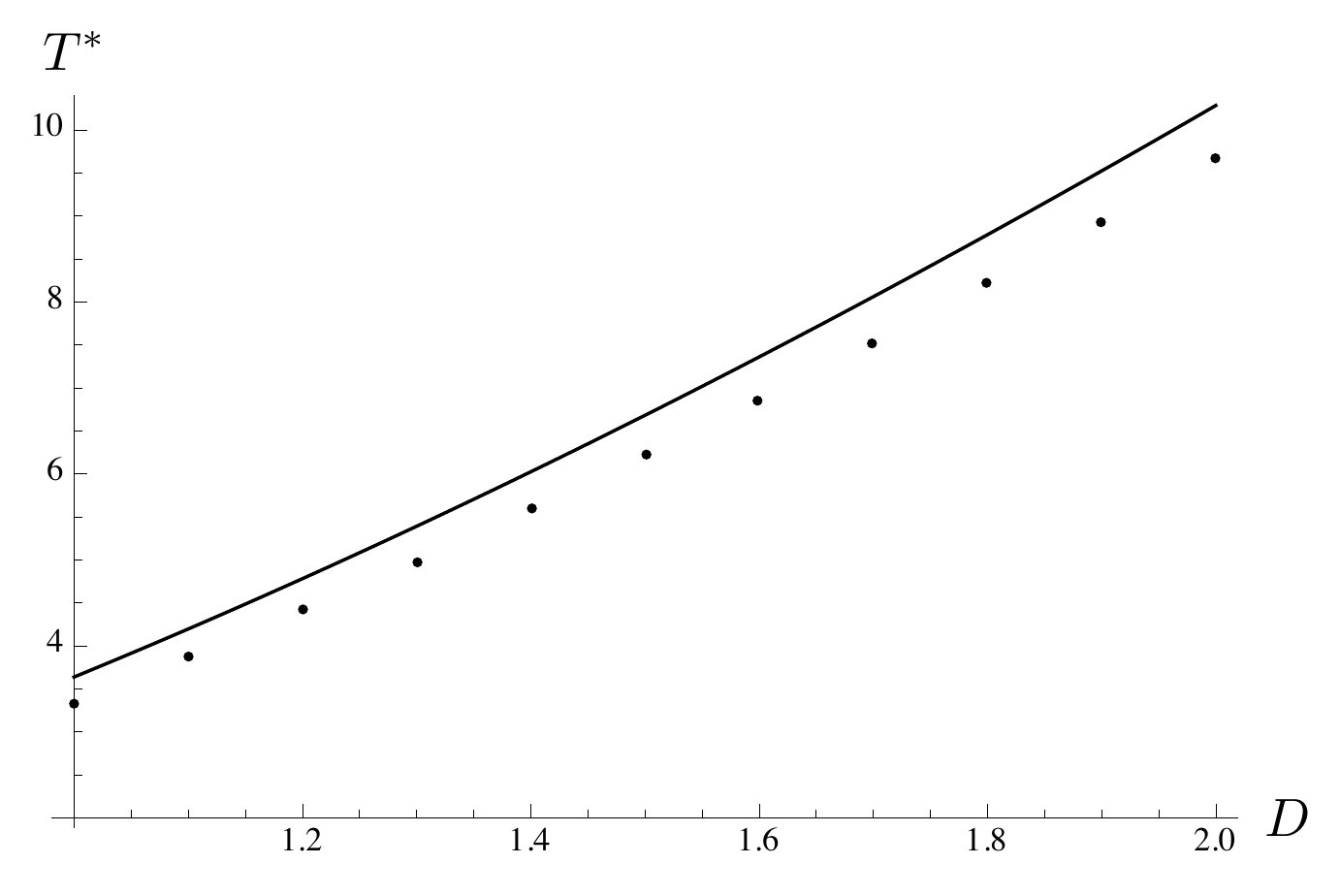} 
   \caption{The value of $T^*$ as a function of $D$ for the relativistic (non-retarded) numerical solution is shown as points -- the parameter $\alpha = \frac{1}{4}$.  The curve is the non relativistic result~\refeq{nonTstar}, again with $\alpha = \frac{1}{4}$.}
   \label{fig:Tstarone}
\end{figure}

We run the iteration~\refeq{noretiter}, and extract from the resulting sequence the value closest to $\half \, D$, returning the time at which it occurs.  We use $\alpha = \frac{1}{4}$ and a time step of $\Delta T = .0005$ to probe a variety of starting displacements $D$, shown in~\reffig{fig:Tstarone}.  There, we can see that the time it takes to halve the initial separation is less than the non-relativistic result -- this is because the force governing the motion is larger than the Coulomb force by a factor of $\frac{1 + |X'|}{1 - |X'|}$ ($X(T)$ is decreasing for $T > 0$, so $X'(T) < 0$).  

\subsection{Retarded Time}\label{ret}

We now return to the full, retarded time description of the problem~\refeq{fullproblemdless} -- all that has changed in the iteration is in the force evaluation -- we now need to solve
\begin{equation}\label{retiter}
\begin{aligned}
X_{j+1} &= X_{j-1} + 2 \, \Delta T \, \frac{P_j}{\sqrt{1 + P_j^2}} \\
P_{j+1} &= P_{j-1}  -\frac{2 \, \alpha \, \Delta T}{ (X_j + X_k)^2} \, \sqof{1 + 2 \, P_k \of{P_k - \sqrt{1 + P_k^2}}} \\
k &= \hbox{min}_\ell \sqof{ |T_j - T_\ell - (X_j + X_\ell)|},
\end{aligned}
\end{equation}
where the last equation defines the index $k$ to be the index $\ell$ that minimizes $|T_j - T_\ell - (X_j + X_\ell)|$, so that $k$ is the appropriate retarded time index, and $T_r \approx k \, \Delta T$.

Operationally, the presence and determination of $k$ is the only major difference between~\refeq{retiter} and~\refeq{noretiter}.  But as a delay differential equation~\refeq{fullproblemdless} requires more ``initial" information.  We must be able to find the correct retarded time for the force evaluation at $T = 0$, and that means we need the past history of our particles.  For now, let's agree that for all time $T< 0$, $X(T) = D$ and $X'(T) = 0$.

\begin{figure}[htbp] 
   \centering
   \includegraphics[width=4in]{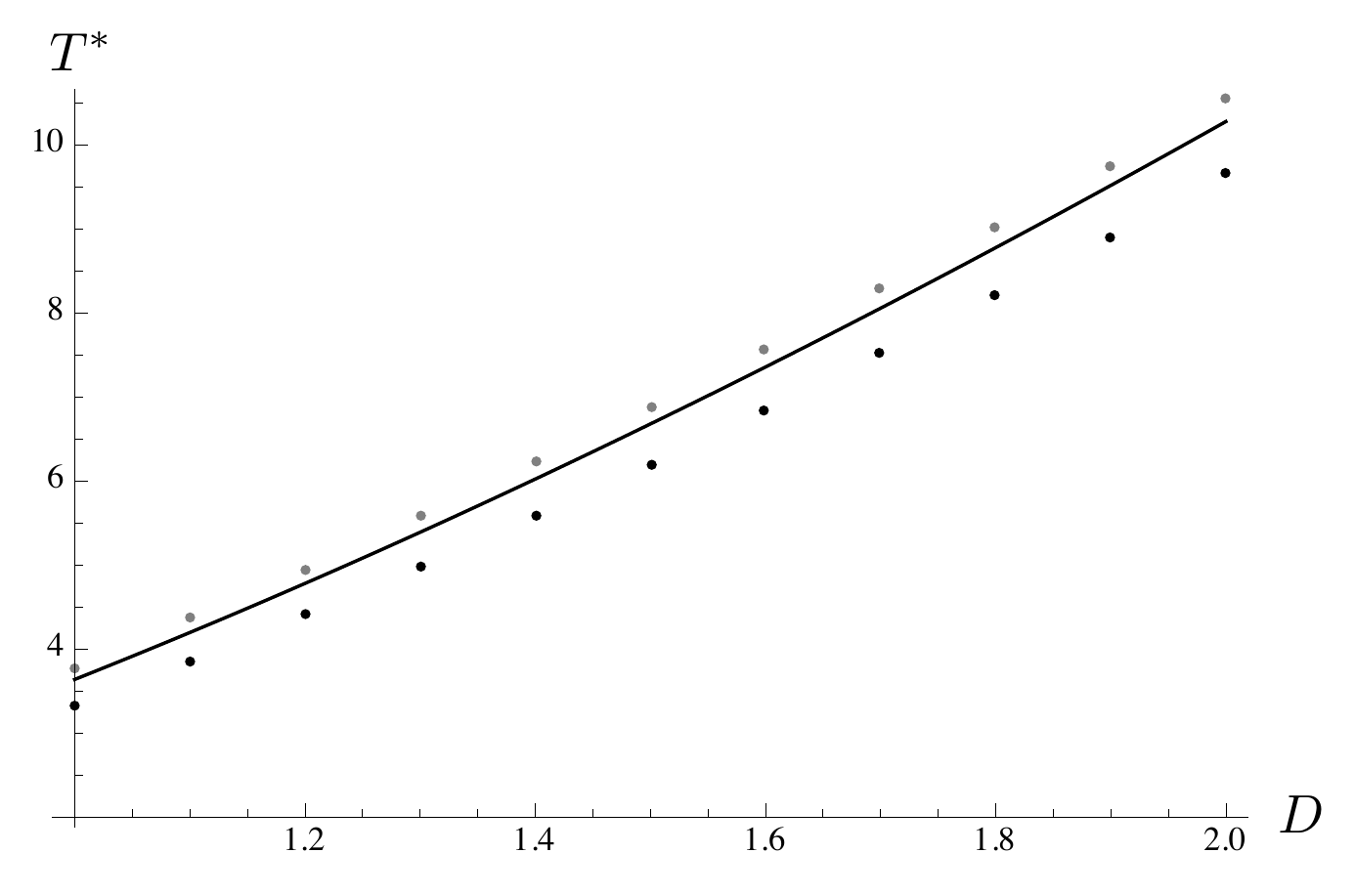} 
   \caption{$T^*$ as a function of $D$ for the relativistic cases with retarded time (top set of points), no retarded time (bottom set of points), and for reference, the non-relativistic result from~\refeq{nonTstar} is shown as a curve.}
   \label{fig:Tstaret}
\end{figure}

A plot of the results, for our test case with $\alpha = \frac{1}{4}$, and a variety of starting values for $D$ is shown in~\reffig{fig:Tstaret}.  The inclusion of retarded time evaluation for the force has an effect here -- the time it takes to reach $\half \, D$ is longer -- that makes sense, since at earlier times, the pair of particles is further apart, and their speed is less.  Then the force on $q$ at $T$ is less using retarded time evaluation than if we use instantaneous evaluation.  It is interesting to note, in~\reffig{fig:Tstaret}, that the $T^*$ values here are better approximated by the non-relativistic, instantaneously evaluated Coulomb force values. 

\section{Synge Iteration}

We are still missing a complete solution to the problem.  In the Newtonian formulation of this physical configuration, we specify that ``at $T = 0$, the pair of particles is at rest with separation $2 \, D$".  If we are to keep that initial value observation, then we technically need to solve~\refeq{retiter} in the absence of trajectory data prior to $T = 0$.  That sounds impossible, and it is hard to imagine how to proceed.  Faced with this ill-posed mathematical question, Synge gave a clever response~\cite{Synge}:  Start with a constant velocity particle trajectory for $-q$ that matches the initial condition at $T = 0$.  Using that, generate the electric fields at $+q$ and allow the positive charge to move under the influence of these fields -- that will give some approximation to the trajectory of the positive charge that is defined for some range $-T_0 < T < T_f$ (where $T_f$ is defined by the initial length of our constant velocity trajectory).  Using that trajectory for $+q$, find the electric field at the negative charge and allow the negative charge to move under its influence -- so we will have an updated trajectory for the motion of $-q$, for a range $-T_1 < T < T_f$.
We lose a piece of the negative charge's trajectory, since the $-T_0$ point of the $+q$ trajectory is not causally connected to the $-T_0$ point of the $-q$ trajectory (it takes some time for the information about $+q$ at $-T_0$ to reach the negative charge, and that means we will have to start at $T_1 < T_0$, where we first have field information at the $-q$ location).  So at each stage of the iteration, the trajectories get shorter.  The process of going back and forth continues iteratively until ``convergence", or until we lose causal contact between the trajectories altogether~\footnote{These trajectories are (numerically) finite, and they shrink as the iteration proceeds, we lose points that are out of causal contact with the current iterative pass.}.

\begin{figure}[htbp] 
   \centering
   \includegraphics[width=4in]{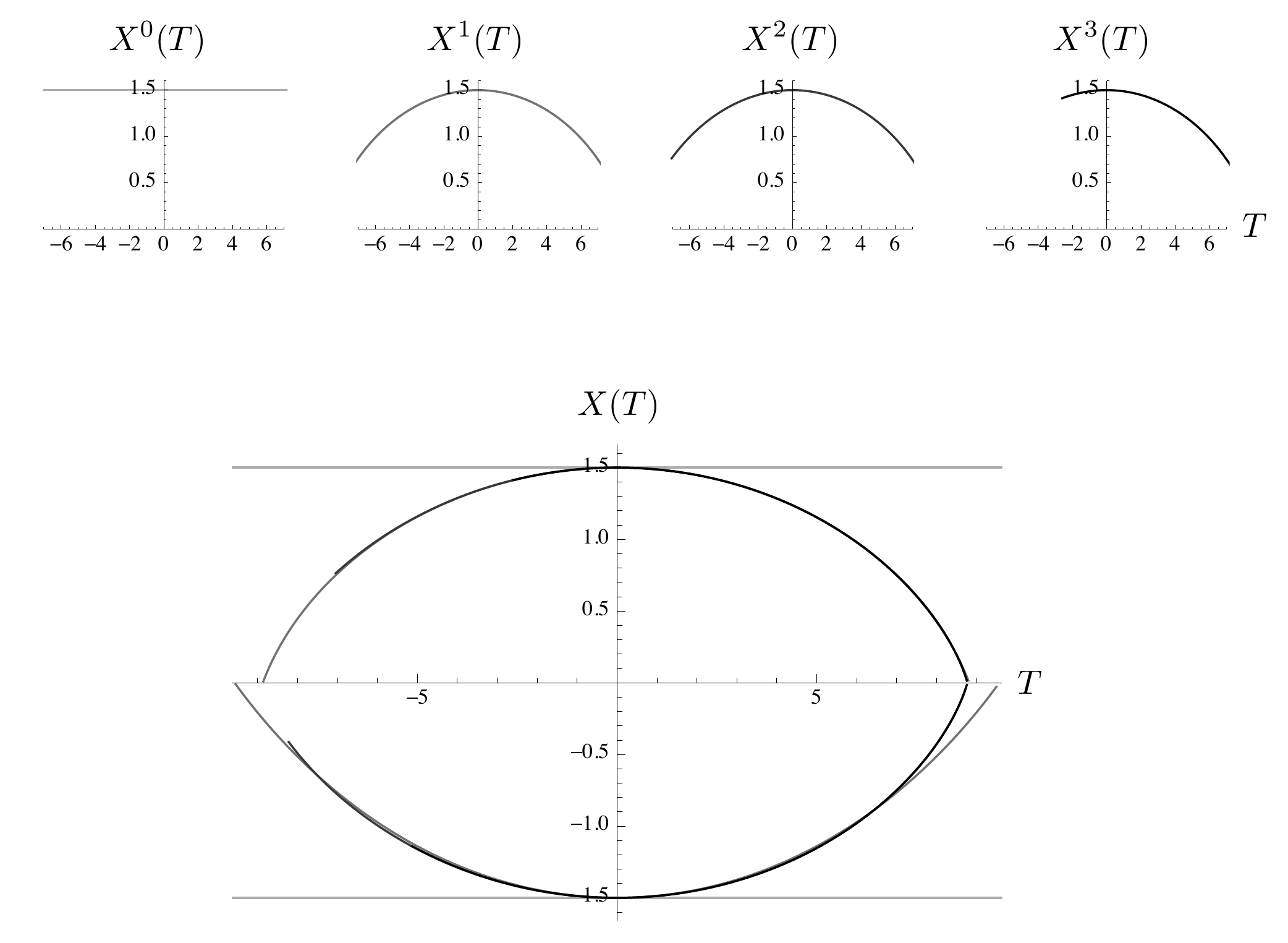} 
   \caption{The top row shows the initial ($X^0(T)$) trajectory, and first three iterates ($X^1(T)$, $X^2(T)$, and $X^3(T)$) for the positive charge as determined by Synge iteration (the two charges start from rest separated by $2 \, D = 3$). The bottom plot shows all four of these on one set of axes, together with the negative charge's trajectories.}
   \label{fig:pair}
\end{figure}

\begin{figure}[htbp] 
   \centering
   \includegraphics[width=4in]{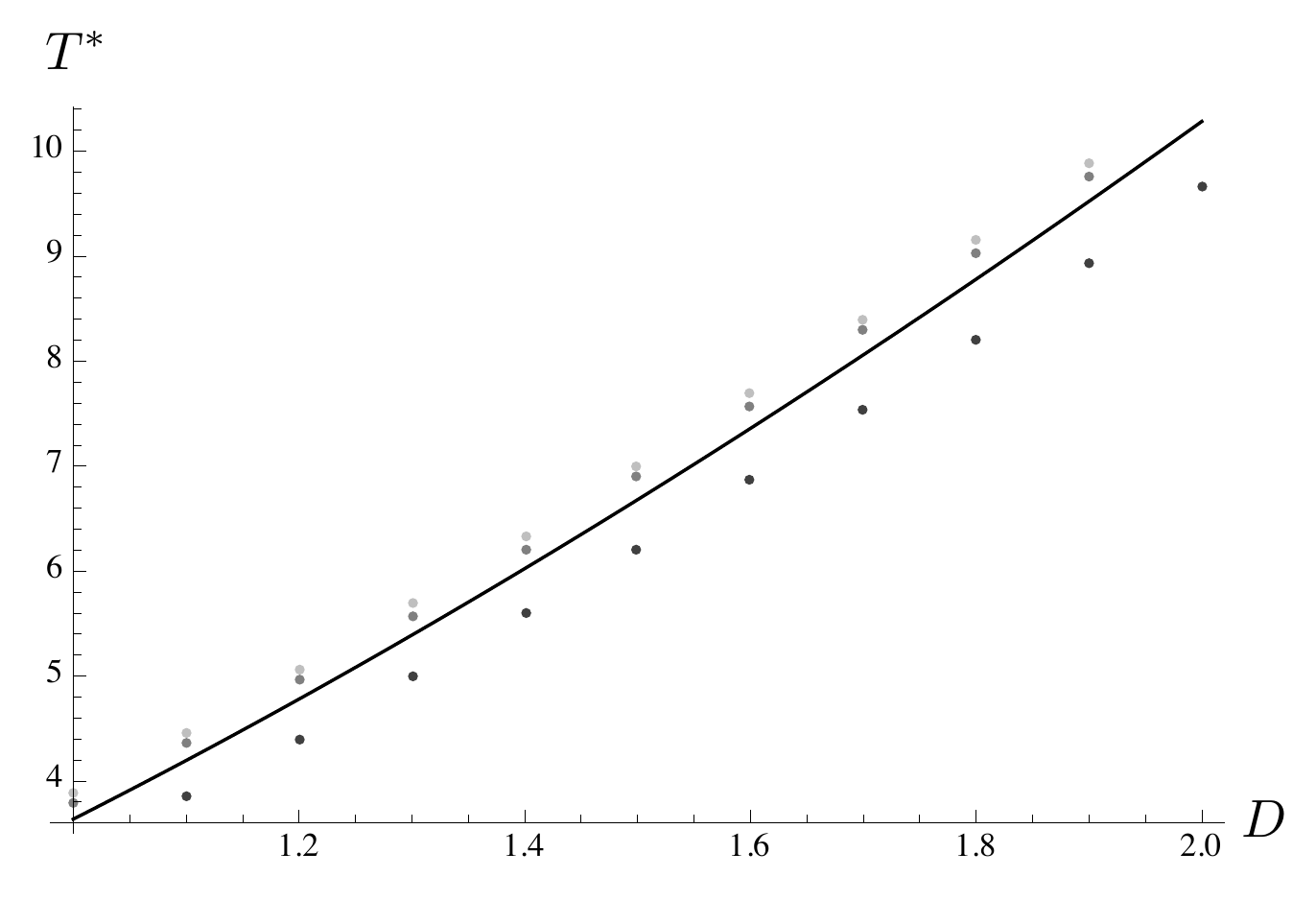} 
   \caption{All four values of $T^*$: fastest (black points) is the relativistic (no retardation) case, then the solid curve is the non relativistic result from~\refeq{nonTstar}, then the relativistic with retardation, and finally, the slowest of all, the Synge iteration considered in this section.  In all cases, we set $\alpha = \frac{1}{4}$. }
   \label{fig:allfour}
\end{figure}

For our one-dimensional problem, the initial trajectory for the positive charge is: ``at rest at $D$", we iterate three times, and the resulting trajectories are shown in~\reffig{fig:pair}.  Note that given this initial trajectory, the first iterate should be precisely the solution trajectory from Section~\ref{ret}.  We can again solve for the time it takes to fall to half the initial position for a variety of initial separations $2\, D$, and once again find a different answer.   For all $D$ we tested, the Synge iteration approach gives a slightly larger value for $T^*$ as shown in~\reffig{fig:allfour}.  

\section{Discussion}

For the non-relativistic and relativistic with no retardation cases, there exist closed form formulae like~\refeq{nonTstar} for comparison (with a numerical method), and those results are relatively straightforward.  Once retardation is introduced in the evaluation of the forces, the standard numerical methods become much more involved.  We have used the simplest possible approach, the sketch~\refeq{retiter} is basically Verlet~\cite{Allen} with a recursive bisection to find the root in~\refeq{fullproblemdless}.  That is just a start when it comes to solving delay differential equations, and we only use it to display the basic qualitative progression shown in~\reffig{fig:allfour}.

The Verlet method we use, defined for the relativistic problem by~\refeq{noretiter}, has per-step accuracy of $O(\Delta T^2)$.  Since we save the position and momentum information on a temporal grid with spacing $\Delta T$, we can only isolate the retarded time to within $O(\Delta T/2)$ (without employing some sort of interpolating function to approximate values off the temporal grid), rendering the method basically $O(\Delta T)$.  For our work here, where we are trying to find the time it takes to get to half the starting separation, we use the rough rule of thumb that we should choose $\Delta T$ so that all of our calculations have:
\begin{equation}
\max{ \| X_c - D/2\| } \le \frac{\Delta T}{2}
\end{equation}
for $X_c$ the position closest to $D/2$ in a given trajectory -- that's what led to $\Delta T = .0005\, $\footnote{The actual bound is much better in most cases, especially in the case of the Synge method, for which we found  $\max{\| X_c - D/2\|} \sim 10^{-5}$}.  While it is clear from~\reffig{fig:allfour} that the four cases we have considered have quantitatively different values for $T^*$, we are not claiming particularly high accuracy for $T^*$.   The accuracy is, however, good enough to distinguish between the cases in~\reffig{fig:allfour} -- the temporal separations between the values of $T^*$ shown there are all well above $\Delta T$.

In terms of physical relevance, taking the Coulomb case as the base level, we see that the non-retarded evaluation of force for the relativistic forces in~\refeq{relforce} is greater than the Coulomb case for the same separation -- this owing to the velocity dependence in the force.  So the acceleration at any separation could easily be greater in the relativistic case, even though the relativistic dynamics puts a cap on the speed (this is opposite the result in~\cite{FMP} where the Coulomb force is used in the relativistic dynamics setting, since there the central body was at rest).  Moving to the relativistic (retarded time) case with fixed position and velocity prior to $T = 0$, we learn that the $T^*$ is much closer to the non-relativistic Coulomb result -- the time it takes to halve the initial distance is longer.  This is because part of the trajectory depends on the position of the particles at $T < 0$, where the separation is fixed at $2 \, D$, larger than the instantaneous evaluation, and that will lead to smaller forces, hence a longer time.  Finally, in the Synge case, the time is again longer -- but there, the $+q$ charge has positive $X'(T_r)$ for $T_r < 0$, which would give a velocity factor $< 1$ in~\refeq{fullproblemdless}, implying a smaller force magnitude than in the relativistic, retarded evaluation case, so again, the force is less, leading to yet more time.

Convergence is hard to define for  the Synge method.  For charges of the same sign, it is not clear that the trajectory of the particles for $T > 0$ depends strongly on the trajectory prior to $T = 0$~\cite{Driver, Hsing} -- that's a surprising result, and suggests that the Synge approach {\it can't} converge in these cases, since the data prior to $T = 0$, which Synge treats in a manner equivalent to the trajectory for $T > 0$, is then basically irrelevant and can be chosen randomly.  Synge, of course, had neither computers nor the almost century of numerical analysis necessary to analyze the convergence and stability of his method in modern terms, and relied on physical insight.  For the cases considered here, the trajectories do indeed appear to converge (that's why it's difficult to pick out the three distinct iterates in the bottom of~\reffig{fig:pair}).  In two dimensions~\cite{Angelov} has shown that the Synge iteration method will not converge to the solution, at least for some ``initial" velocities (those specified at $t = 0$).

\section{Conclusion}

The problem we discuss here, that of the self-consistent motion of a pair of charged particles, has well-known intricacies and has been studied carefully for over a hundred years.  We revisit this interesting configuration to remind ourselves that while classical E\&M is the most complete and tested of the classical forces of nature, there remain mysteries that defy solution.  This problem of motion is also an interesting vehicle for {\it introducing} new physics in the context of a familiar, and easy to state, problem.

In~\reffig{fig:allfour}, we offer four physically distinct solutions to the problem of electromagnetic infall, each probing a different regime -- where is the final, full case, solved by Synge iteration, physically relevant?  For consistency with the relativistic point fields, we use the relativistic equations of motion, and that means that the particles we are interested in are moving at relativistic speeds.  In addition, we would need particles that are widely separated, so that the time-of-flight for the force information was significant (i.e.\ so that we must use the full retarded force evaluation).  Finally, we need particles whose past histories are unknown, not ones that have been sitting in an electron-gun waiting to be fired -- in a sense, then, we need particles that are free of external interaction.

Aside from physical interest in these astronomically separated massless charged particles moving very quickly, the Synge method provided an approach to solving problems in which past histories are not known.  In two dimensions,  the Synge iteration technique does not converge to solutions of the Synge problem (when they exist)~\cite{Angelov}.  In one dimension, we have side-stepped the issue of existence and convergence for our problem, relying instead on the ``experimental" verification provided by~\reffig{fig:pair} (and previous work on the one-dimensional problem,~\cite{Hsing}, for example).  But even in this simplified setting, there are known physical difficulties:  notably, there is a lack of convergence implied by~\cite{DriverOneD, DriverCan} in which particles of like charge have $t > 0$ trajectories that are fixed, in some cases, by specifying only $x(0)$ and $\dot x(0)$.  How does the Synge method handle the lack of dependence on the $t < 0$ portion of the trajectory?   In addition to this question, it would be interesting to introduce the radiation reaction self-force in~\refeq{reom} -- this force depends on $t$ rather than $t_r$, and its effects could be studied in either the full ``unknown" past history case (from Section IV), or in the artificial ``particles at rest for all $t < 0$" limit (from Section III B).

%
%
%

\begin{acknowledgements}
The authors thank Professor David Griffiths for useful commentary and suggestions.
\end{acknowledgements}

\end{document}